\documentclass[aps,prl,twocolumn,showpacs]{revtex4}
\usepackage[T1]{fontenc}
\usepackage[latin1]{inputenc}
\usepackage[dvips,pdftex]{graphicx}
\usepackage{hyperref}
\usepackage{amsmath}
\usepackage{ulem}



\begin{document}
\title{Anderson localization or nonlinear waves? A matter of probability}
\author{M.V.~Ivanchenko$^{1,2}$, T.V.~Laptyeva$^2$ and S.~Flach$^2$}

\affiliation{$^1$ Theory of Oscillations Department, University of Nizhniy Novgorod, Russia
\\ $^2$ Max-Planck-Institut f\"ur Physik komplexer Systeme, N\"othnitzer Str. 38, 01187 Dresden, Germany}

\begin{abstract}
In linear disordered systems Anderson localization makes any wave packet stay localized for all times. Its fate in nonlinear disordered systems is under intense theoretical debate and experimental study. We resolve this dispute showing that at any small but finite nonlinearity (energy) value there is a finite probability for Anderson localization to break up and propagating nonlinear waves to take over. It increases with nonlinearity (energy) and reaches unity at a certain threshold, determined by the initial wave packet size. Moreover, the spreading probability stays finite also in the limit of infinite packet size at fixed total energy. These results are generalized to higher dimensions as well.
\end{abstract}

\maketitle

Anderson localization (AL) is a fundamental physical wave phenomenon with exponential localization of eigenstates in linear wave equations due to disorder \cite{Kramers,Mirlin}. 
It was originally used to explain metal-insulator transitions \cite{Anderson}, and has been recently related to experimentally observed localization of weak intensity light \cite{Optics} and noninteracting matter waves \cite{BEC} on disordered lattice systems. The fate of AL in presence of many body interactions or corresponding nonlinearity has stayed at frontiers of theoretical and experimental physics since \cite{nonlinear1,nonlinear2}.

The pioneering experiments \cite{lucioni} demonstrate both the principal possibility to study this effect, but also the severe length and time scale 
limitations which restrict a quantitative analysis. Theoretical studies become therefore central here. Two faces of Anderson localization are usually probed: the absence of the wavepacket spreading \cite{molina,Pikovsky_Shepelyansky,Flach09,Pikovsky,Mulansky,skokos1,skokos2,Flach10,laptyeva10,bodyfelt11} (the focus of this paper) and vanishing conductivity \cite{basko,ourpramanapaper}. Wave packets, for not too weak nonlinearities, have been found to spread subdiffusively, disobeying AL at least over many time scales \cite{molina,Pikovsky_Shepelyansky,Flach09}, and remarkable progress in studying the regimes and characteristics of wave packet spreading has been achieved \cite{skokos1,skokos2,Flach10,laptyeva10,bodyfelt11}. 

Still, the original questions are open. It remains debated, whether the observed spreading will continue infinitely or slow down and even stop to restore AL, once the wave packet densities become substantially small, as conjectured in Ref. \cite{Pikovsky,Mulansky,Magnus}. It is unknown, whether there exists a lower bound on the nonlinearity strength, beyond which wave packets do obey AL, and therefore do not spread at all; or are some time scales diverging in this limit? Unavoidable finite size, time, energy, and precision limitations will make the most advanced numerical results not entirely convincing. To achieve a breakthrough, a novel approach is in need. 

The probabilistic description of AL in terms of the measure of localized regular trajectories in phase space (periodic orbits and tori) appears a promising solution for nonlinear disordered systems. Indeed, the mere assumption that a wave packet is launched on a chaotic trajectory leads to the conclusion that chaos remains forever:  
Arnol'd conjecture, unproved but widely accepted, states the uniqueness of the chaotic region in phase space \cite{Arnold}. If initial conditions are chaotic, then the trajectory will be unbounded in phase space, corresponding to unlimited spreading of a wavepacket. If they belong to a localized periodic orbit or torus, then the spreading is absent. 

The progress has been quite limited up to now due to the difficulty of the task. Persistence of tori has been proved for a special class of infinite systems \cite{FSW} and for finite tori dimensionality \cite{Wang_finite} only. The non-zero probability of existence of infinite-dimensional tori in generic case (for small enough energies) has been conjectured by empirical and numerical arguments in Ref. \cite{Magnus}. Quantitative estimates for AL probabilities are lacking. 

In this paper we show the probabilistic nature of AL in nonlinear disordered systems. At any small but finite nonlinearity (energy) value there is a finite probability for AL to break up and for nonlinear waves to propagate. It increases with nonlinearity (energy) and reaches unity at a certain threshold, determined by the initial wave packet size. Moreover, the spreading probability stays finite also in the limit of infinite packet size at fixed total energy. This finite value is between zero and one for quartic anharmoncity (as realized for two body interactions and in optical Kerr media), strictly one for the lower degree of anharmonicity, and has zero as the lower bound for higher degrees of anharmonicity.



We first address the paradigmatic Fr\"ohlich-Spencer-Wayne (FSW) type $d$-dimensional classical lattice \cite{FSW}, for which the linear eigenmodes of a disordered system are compact single-site excitations with 
random frequencies, and where infinite dimensional Kolmogorov-Arnold-Moser (KAM) tori \cite{FSW} exist.
This model can be considered as the strong disorder limit of generic classical Klein-Gordon (KG)
and semi-classical discrete nonlinear Schr\"odinger 
(DNLS) arrays, where the existence of KAM tori is still debated \cite{Wang_finite,Magnus}.
Its Hamiltonian reads
\begin{equation}
\label{eq1}
\begin{aligned}
& \mathcal{H}=\frac{1}{2}\sum\limits_{\boldsymbol{n}}\left[p_{\boldsymbol{n}}^2+\varepsilon_{\boldsymbol{n}}x_{\boldsymbol{n}}^2
+\frac{1}{\gamma}\sum\limits_{\boldsymbol{m}\in D(\boldsymbol{n})}
(x_{\boldsymbol{m}}-x_{\boldsymbol{n}})^\gamma\right],
\end{aligned}
\end{equation}
where $x_{\boldsymbol{n}}$ is the displacement of the
$\boldsymbol{n}=(n_1,\ldots,n_d)$-th particle from its original
position, $p_{\boldsymbol{n}}$ its momentum,
$D(\boldsymbol{n})$ is the set of its nearest neighbors, and the random uncorrelated $\varepsilon_n\in[1/2,3/2]$ are uniformly distributed. Unless explicitly specified, a chain $d=1$ with 
quartic anharmonicity $\gamma=4$ is considered. Without the loss of generality we assume $\varepsilon_0=1$.

We consider a wave packet with initial width $L$ and determine the probability of $L$-site localized periodic and quasi-periodic solutions of (\ref{eq1}). 
We start with periodic orbits, which are {\it single-site localized} solutions, and derive conditions of their destruction (when regular trajectories delocalize the 
corresponding wavepackets are assumed to spread). Let us construct an exact time-periodic orbit of (\ref{eq1}) localized at $n=0$. 
We apply perturbation theory $x_n(t)=\sum_{k=0}^\infty x_n^{(k)}(t)$ 
in the small-amplitude limit, taking $x_n^{(0)}(t)=A_n\cos(t)\delta_{0,n}$ as the zero-order approximation (note that $\varepsilon_0=1$). In first order we find
\begin{equation}
\label{eq2_0}
x_{\pm 1}(t)=A_{\pm 1}\cos{t}, \ A_{\pm 1}=\frac{3 A_{0}^3}{4(\varepsilon_{\pm 1}-\varepsilon_0)}.
\end{equation}
In higher orders it follows
\begin{equation}
\label{eq2}
x_{\pm n}(t)=A_{\pm n}\cos{t}, \ A_{\pm n}=\frac{3 A_{\pm (n-1)}^3}{4(\varepsilon_{\pm n}-\varepsilon_0)}.
\end{equation}
For the perturbative solution (\ref{eq2}) to converge the amplitudes must decay: 
\begin{equation}
\label{eq3}
\left|\frac{A_{\pm n}}{A_{\pm (n-1)}}\right|=\frac{3 A_{\pm (n-1)}^2}{4|\varepsilon_{\pm n}-\varepsilon_0|}<\frac{1}{\kappa}, \ \kappa>1.
\end{equation}
The probability for this condition to hold at site $\pm n$ is determined by its random on-site potential $\varepsilon_{\pm n}$ and the oscillation amplitude of its neighbor: 
\begin{equation}
\label{eq4}
\mathcal{P}^{(\pm n)}=1-\frac{3}{2}\kappa A_{\pm (n-1)}^2\ge 1-\frac{3}{2} \kappa^{3-2n} A_{0}^2.
\end{equation} 
The probability to obtain a localized time-periodic solution in the infinite chain $\mathcal{P}=\prod_{n=1}^\infty \left[\mathcal{P}^{(n)}\right]^2$ is bounded from above
by the probability to have decreasing amplitudes at least in the first neighbors
\begin{equation}
\label{eq5}
\mathcal{P}\le \mathcal{P}^{(1)}\mathcal{P}^{(-1)}=\left(1-3\kappa E\right)^2 \equiv \mathcal{P}^+,
\end{equation}
and from below by the probability to get at least an exponentially decaying profile in the infinite system
\begin{equation}
\label{eq6}
\mathcal{P}\ge\prod\limits_{n=1}^\infty\left( 1-\frac{3}{2} \kappa^{3-2n} A_{0}^2\right)^2
\ge \left(1-\frac{3\kappa E}{1-\kappa^{-2}}\right)^2 \equiv \mathcal{P}^-,
\end{equation}  
where $3\kappa E\ll 1$ was used in (\ref{eq6}) and where $E=A_{0}^2/2$ is the energy of the central site.

It follows from (\ref{eq5}) that it is not possible to construct a {\it single-site localized} ($|A_n/A_0|\ll1, \ \forall n\neq 0$) time-periodic orbit for $E>1/3$. 
For smaller energies, the lower 
bound of spreading probability always remains non-zero and scales linearily with the total wavepacket energy $E$, backing the arguments in \cite{Magnus}.

Next we consider $L$-site localized solutions to (\ref{eq1}) loosely corresponding to tori. Their existence probability is maximized
for sparse packets, when the most excited sites are separated by intervals of weakly excited ones (due to reduced perturbative corrections, or, equivalently, 
resonance probabilities, see also \cite{Magnus}). We assume such a sparse excitation with $L$ sites with the energy $E/L$ per site, separated by at least two non-excited sites, 
as a zero order approximation for the perturbation theory. In the first order the problem is reduced to $L$ independent single-site problems and, with (\ref{eq5}), the upper 
bound for the localization probability reads
\begin{equation}
\label{eq7}
\mathcal{P}_L=\left(1-\frac{3\kappa E}{L}\right)^{2L}.
\end{equation}
Note, that one can proceed to arbitrary high orders assuming sparser zero-order approximations, ultimately obtaining the lower bound analogue to (\ref{eq7}) by exponentiating (\ref{eq6}). 
$\mathcal{P}_L$ is a monotonously increasing function of $L$ with the limiting localization probability for the infinite-size packet 
\begin{equation}
\label{eq8}
\mathcal{P}_\infty=e^{-6\kappa E}.
\end{equation}
Eqs. (\ref{eq7}) and (\ref{eq8}) are central results of this work. 
Note, that a wave packet of size $L$ has $2L$ phase space variables to be defined. Therefore, its relevant phase space dimension is $2L$.
The probabilitiy of obtaining regular AL states (or not) is defined through the ratio of the volume $v_l$ of all points in this $2L$-dimensional phase space,
which yield AL localization, to the full available volume $v_l+v_s$ where $v_s$ is the volume of all points which yield spreading: $\mathcal{P}_L = v_l/(v_l+v_s)$.
We can now conclude, that
for a wave packet of size $L$ no regular AL states are expected if the energy density $h\equiv E/L > 1/3$.
But even for $h < 1/3$ there is always a finite probability to observe spreading trajectories. Most interesting is that given a fixed total energy $E$, the probability
for AL is approaching a finite value below one in the limit  of infinite packet size $L$ (\ref{eq8}). 
Therefore even in this limit (of zero energy density and infinite packet size) there remains always a nonzero probability
to spread, i.e. $v_s/(v_l+v_s) \neq 0$ in this limit! 
Remarkably also the derivative $\partial \mathcal{P}_L / \partial L \sim L^{-2}$ is vanishing as a power law, and not as an exponential for large $L$. Therefore, at variance to the
case of AL, there is no new length scale emerging. In particular, already the first moment $\langle L \rangle$ obtained with such a probability distribution function diverges.

%


\begin{figure}[t]
{\centering
\resizebox*{0.95\columnwidth}{!}{\includegraphics{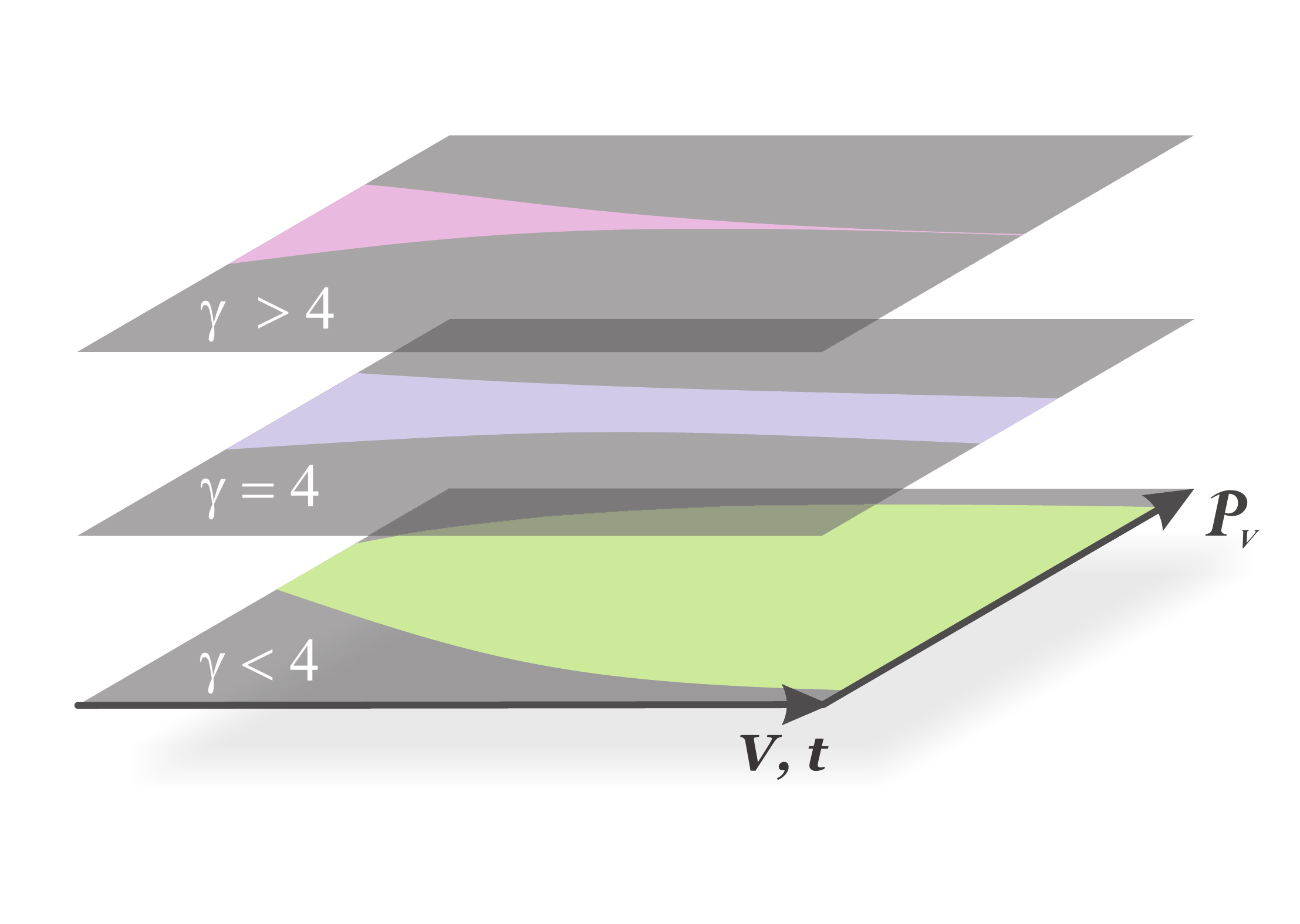}}}
{\caption{(Color online) Schematic dependence of the probability $\mathcal{P}_V$ for wave packets to stay localized (dark area) together with the complementary
light area of spreading wave packets versus the wave packet volume $V$ (either initial or attained at some time $t$) for three different orders of nonlinearity
$\gamma<4$, $\gamma=4$ and $\gamma > 4$.
}\label{fig3}}
\end{figure}

We generalize our results to the case of arbitrary degree of nonlinearity $\gamma$ and lattice dimensionality $d$.
We consider localized solutions as a sparse wave packet of $V$ equally excited sites and characteristic size $L$ and derive the upper bound for their existence 
probability $\mathcal{P}_V$. As its volume scales as $V\propto L^d$ we straightfowardly arrive at the probability to find an AL wave packet
\begin{equation}
\label{eq9}
\begin{aligned}
&\mathcal{P}_V=\left(1- \frac{\kappa_\gamma E^{\gamma/2-1}}{V^{\gamma/2-1}}\right)^{2 V d}.
\end{aligned}
\end{equation}
We find that for any $\gamma$ no regular AL wave packets are expected if the energy density $h > h_\gamma$. For smaller $h$ there is always a finite probability to launch a spreading
wave packet. Most interesting is that the fraction of localized AL wave packets $v_l/(v_l+v_s)$ tends to zero in the limit $V \rightarrow \infty$ at fixed $E$ for $\gamma < 4$, and tends to unity for $\gamma > 4$ (note, that this is only the upper bound for the AL probability), as shown schematically in Fig.\ref{fig3}.


\begin{figure}[t]
{\centering
\resizebox*{0.95\columnwidth}{!}{\includegraphics{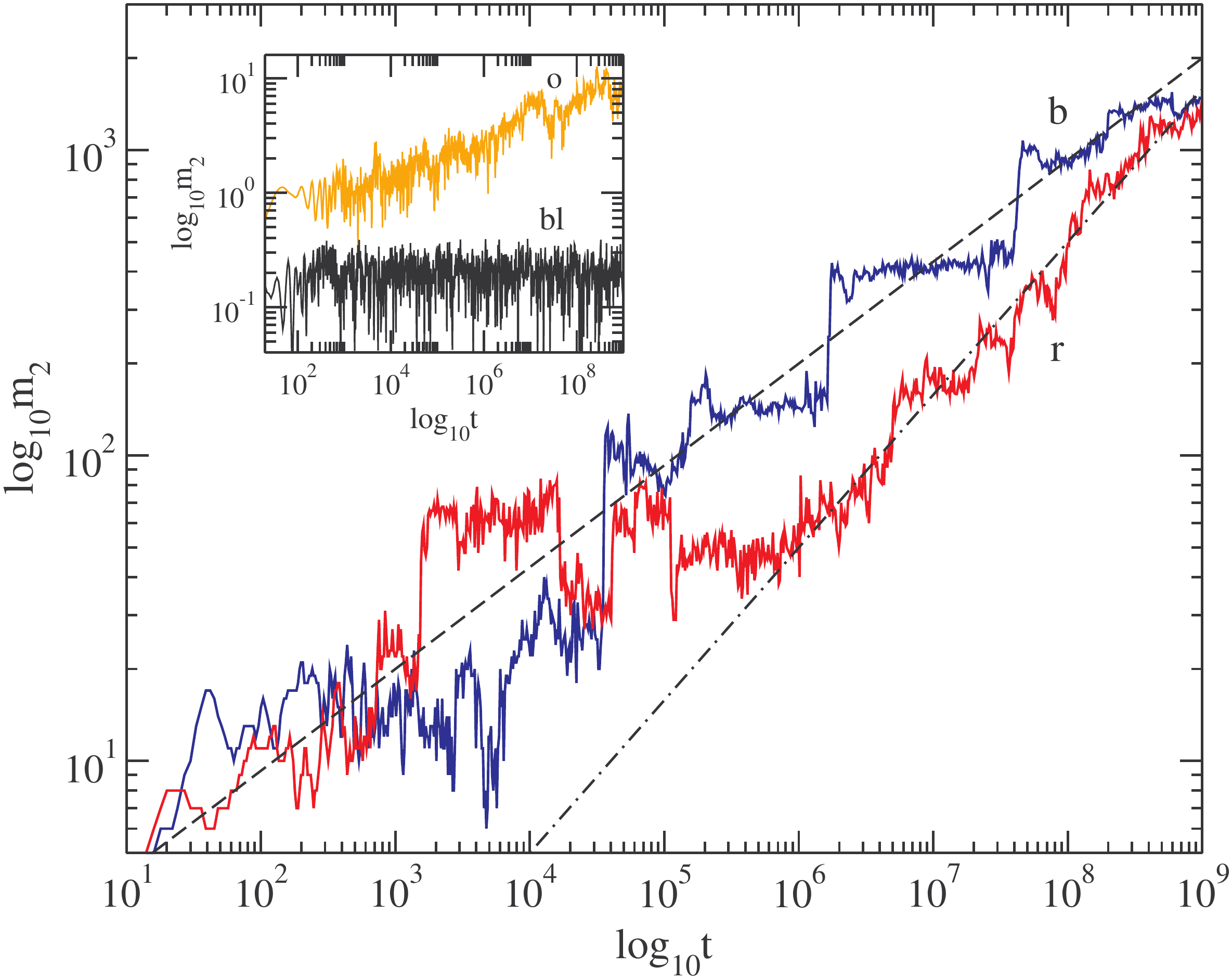}}}
{\caption{(Color online)  Second moment of  a spreading single site excitation for two different disorder realizations versus time, $d=1$, $\gamma=4$, $E=1$. Dashed and dash-dotted lines with the slopes $1/3$ and $1/2$, respectively, guide an eye. 
Inset: Second moment of a spreading and a nonspreading single site excitation for two different disorder realizations versus time, $d=1$, $\gamma=4$, $E=0.05$.
}\label{fig1}}
\end{figure}

To probe our analytical results we simulate the evolution of single-site initial excitations with energy $E$ in (\ref{eq1}) and (\ref{KGeq1}) for $d=1, \ \gamma=4$. 
We use a symplectic SABA-type scheme \cite{skokos1,bodyfelt11} up to final times $t_{end}=10^9$. To characterize the wave packet we calculate its 
second moment $m_2=\sum\limits_n (n-\bar{n})^2 E_n/E$, with $\bar{n}=\sum\limits_n n E_n/E$ which measures the squared distance between both packet tails, and the
participation number $P=\frac{E^2}{\sum\limits_n E_n^2}$ which tells the number of most strongly excited lattice sites.

We first choose an energy $E=1$ which according to our above results should yield spreading wave packets with probability one.
Indeed in Fig.\ref{fig1} we plot the growth of $m_2$ for two disorder realizations, which survives a fit $m_2\propto t^{1/3\ldots 1/2}$ in accord with
theoretical preditions \cite{Flach10,laptyeva10}.
For small energy $E=0.05$ we expect to observe both spreading and nonspreading wave packets. In the inset in Fig.\ref{fig1} we show two typical cases -
in one case the second moment does grow, in the second case it does not, therefore indicating AL.


\begin{figure}[t]
{\centering
\resizebox*{0.95\columnwidth}{!}{\includegraphics{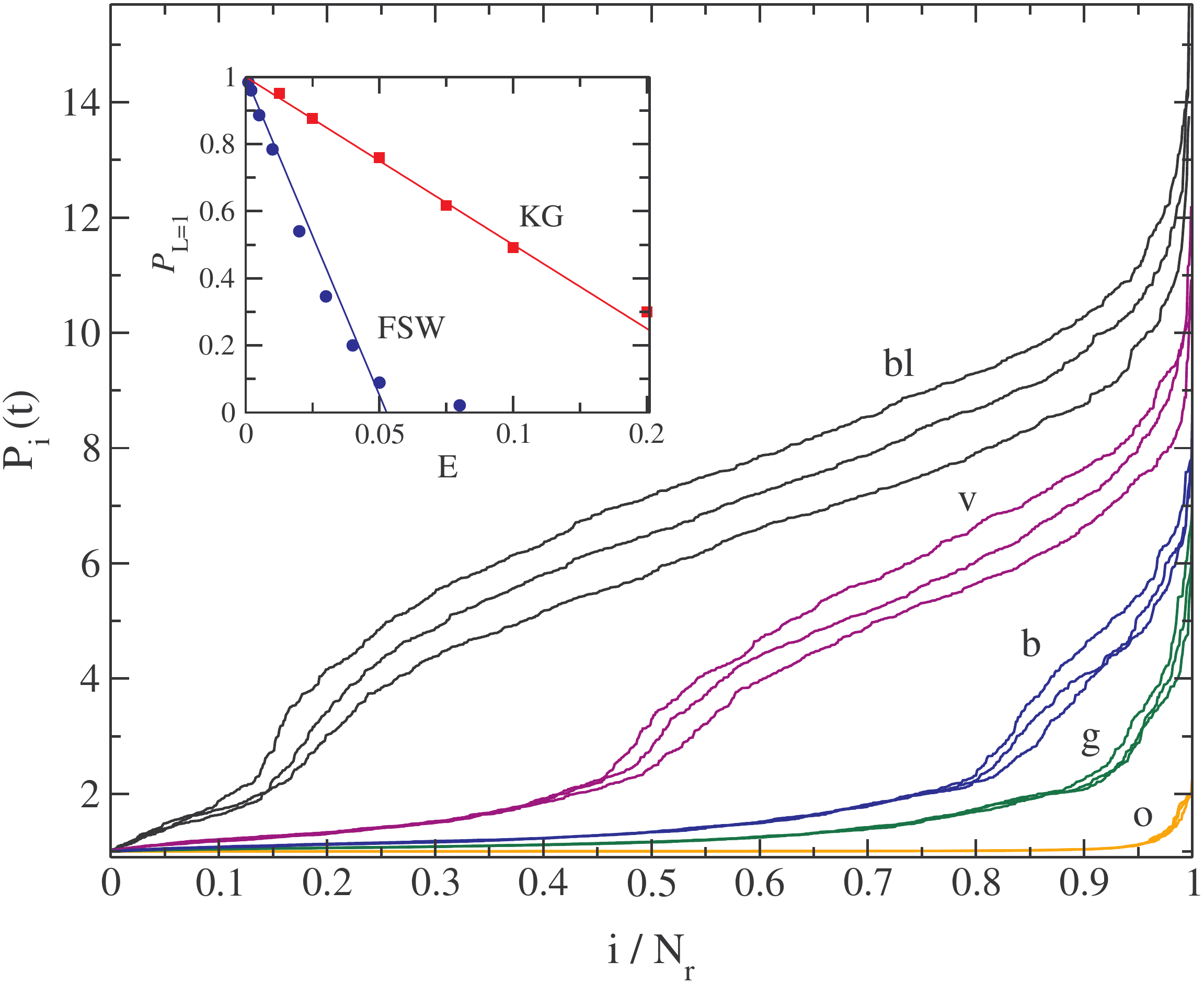}}}
{\caption{(Color online) Sorted participation numbers $P_i$ for $N_r=1000$ different disorder realizations $1 \leq i \leq N_r$ at $t=10^8, 5\cdot 10^8, 10^9$ (bottom-top) for $E=0.08$ (bl), 
$E=0.05$ (v), $E=0.03$ (b), $E=0.02$ (g), $E=0.002$ (o). Inset: single-site localized fractions ($\mathcal{P}_{L=1}$): numerics by symbols, linear fits  
for FSW and for KG by solid lines.
}\label{fig2}}
\end{figure}
In order to quantify the above results, we perform simulations of single site excitations at a given energy $E$ for $N_r=1000$ different disorder realizations.
We measure the participation number $P(t)$ at three large time values $t=10^8, 5\cdot 10^8, 10^9$. For each time we sort the participation numbers by their value
and plot them in Fig.\ref{fig2} versus the sorting index $i$. We find that for very small energy $E=0.002$ all three curves for different times overlap and have small values of $P$,
indicating AL at least up to the largest time. For larger energies, e.g. $E=0.05$ we find that the curves overlap up to some largest index $i_m(E)$, and differ for larger indices.


To quantify the results we introduce the following numerical criterion for a single-site (or a single-mode) localized trajectory: the one with participation number $P(t)<1+\alpha,$ choosing a small enough $\alpha=0.2$. As $P\approx1+2 (E_{-1}+E_{1})/E_{0}$ for strongly localized trajectories, (\ref{eq4}) yields a theoretical estimate for the non-spreading fraction: $ \mathcal{P}\approx1-6\sqrt{2/\alpha}E$.
The numerically obtained dependence of the non-spreading fraction on the energy shows a linear decay and agrees well with this analytical estimate (Fig.\ref{fig2}, inset).

A similar approach can be developed the Klein-Gordon (KG) chain described by the Hamiltonian
\begin{equation}
\label{KGeq1}
\begin{aligned}
& \mathcal{H}=\frac{1}{2}\sum\limits_{\boldsymbol{n}}\left[p_{\boldsymbol{n}}^2+\varepsilon_{\boldsymbol{n}}x_{\boldsymbol{n}}^2
+\frac{u}{4}x_{\boldsymbol{n}}^4+\frac{1}{2W}\sum\limits_{\boldsymbol{m}\in D(\boldsymbol{n})}
(x_{\boldsymbol{m}}-x_{\boldsymbol{n}})^2\right],
\end{aligned}
\end{equation}
This is a generic model with a finite localization length of Anderson modes \cite{Flach09,skokos1,skokos2,laptyeva10,bodyfelt11}. 
We performed a similar numerical analysis at $u=1$ as for the FSW model with launching a single eigenmode of the corresponding linear eigenvalue problem of $u=0$.
The parameter $W=6$.
The final outcome is plotted in the inset in Fig.\ref{fig2}. it shows again a linear decay of $\mathcal{P}_L$ with energy, similar to the FSW model.

In summary, we developed the theory of AL in nonlinear disordered systems in zero-temperature limit. In contrast to linear systems, where AL forces any wave packet to stay localized for all times, the fate of AL becomes a matter of probability and nonlinear waves may propagate. With increasing the strength of nonlinearity (energy) the probability of AL is reduced, 
and reaches zero at a certain level of nonlinearity depending on the initial wave packet size. At fixed total energy, the fraction of AL wave packets does not reach unity in the limit of an infinite packet size, but stays finite yet less than one for quartic anharmonicity. For weaker degree of nonlinearity the AL fraction in this limit
vanishes completely, while for stronger nonlinearity degree its upper bound tends to one. It follows that the previously conjectured slowing down or halt of wave propagation \cite{Pikovsky,Mulansky,Magnus} are not realized at least for the quartic and weaker anharmonicity. The observation of a finite fraction of spreading wave packets at small energies questions the correctness of the perturbation calculations in \cite{dynamical}, which seem to exclude such a possibility. These results generalize to higher dimensions as well. An analytic treatment of the generic KG-type system yields similar results and will be presented in more detail elsewhere. Our results not only resolve a fundamental theoretical problem in nonlinear wave physics but also have multifaceted experimental implications, as for design, as for 
interpretation.

MI acknowledges financial support of the Dynasty Foundation.


\end{document}